\documentclass{iopart}
\usepackage[T1]{fontenc}
\usepackage[latin9]{inputenc}
\usepackage{geometry}
\usepackage{units}
\usepackage{amsbsy}
\usepackage{graphicx}
\usepackage{setspace}
\usepackage{esint}
\makeatletter

\newcommand{\lyxdot}{.}

\usepackage{iopams}
\usepackage{setstack}

\usepackage{cite}

\usepackage{ifthen}\renewenvironment{figure}[1][]{%
 \ifthenelse{\equal{#1}{}}{%
   \@float{figure}
 }{%
   \@float{figure}[#1]%
 }%
 \centering
}{%
 \end@float
}

\makeatother

\begin{document}

\title{Singularities in Large Deviation Functionals of Bulk-Driven Transport
Models}

\author{Avi Aminov$^{1}$, Guy Bunin$^{2}$ and Yariv Kafri$^{1}$}

\address{$^{1}$ Faculty of Physics, Technion - Israel Institute of Technology,
Haifa 32000, Israel}

\address{$^{2}$ Department of Physics, Massachusetts Institute of Technology,
Cambridge, MA 02139, USA}

\ead{aviamino@tx.technion.ac.il, buning@mit.edu, kafri@physics.technion.ac.il}

\begin{abstract}
The large deviation functional of the density field in the weakly
asymmetric simple exclusion process with open boundaries is studied
using a combination of numerical and analytical methods. For appropriate
boundary conditions and bulk drives the functional becomes non-differentiable.
This happens at configurations where instead of a single history,
several distinct histories of equal weight dominate their dynamical
evolution. As we show, the structure of the singularities can be rather
rich. We identify numerically analogues in configuration space of
first order phase transition lines ending at a critical point and
analogues of tricritical points. First order lines terminating at
a critical point appear when there are configurations whose dynamical
evolution is controlled by two distinct histories with equal weight.
Tricritical point analogues emerge when there are configurations whose
dynamical evolution is controlled by three distinct histories with
equal weight. A numerical analysis suggests that the structure of
the singularities can be described by a Landau like theory. Finally,
in the limit of an infinite bulk bias we identify singularities which
arise from a competition of $s$ histories, with $s$ arbitrary. In
this case we show that all the singularities can be described by a
Landau like theory. 
\end{abstract}
\maketitle

\section{Introduction}

In recent years there has been much focus on understanding full probability
distributions in non-equilibrium systems. In particular, much progress
has been achieved in the context of driven diffusive systems \cite{derrida2007non}.
For these systems there is a growing understanding of both probability
distributions of currents \cite{bertini2005current,bodineau2005distribution,lecomte2009current,merhav2010bose,krapivsky2012fluctuations,gorissen2012exact,meerson2013extreme,akkermans2013universal}
and density profiles \cite{derrida1993exact,derrida2002large,derrida2003exact,enaud2004large,bertini2010lagrangian,bunin2012large,cohen2011phase,cohen2014nonequilibrium}.\textbf{
}In the latter case, which is the focus of this paper, it can be shown
that for a large class of systems the probability distribution of
a density field $\rho({\bf x})$ obeys a large deviation principle
\cite{kipnis1989hydrodynamics,touchette2013large}: 
\begin{equation}
P\left[\rho({\bf x})\right]\sim e^{-N\phi\left[\rho({\bf x})\right]}\;.
\end{equation}
Here ${\bf x}$ is a spatial coordinate, $N$ is the system size,
$P\left[\rho({\bf x})\right]$ is the probability functional of the
density field and $\phi\left[\rho({\bf x})\right]$ is the large deviation
functional (LDF). In equilibrium $\phi\left[\rho({\bf x})\right]$
is given by the free energy of the system. Therefore, out of equilibrium
it can be considered as a direct analogue of the free energy.

In equilibrium, when the system is a diffusive gas in a disordered
phase and the interactions are short ranged $\phi\left[\rho({\bf x})\right]$
is a local and smooth functional. By smooth it is meant that as $\rho({\bf x})$
is varied smoothly the functional $\phi\left[\rho({\bf x})\right]$
changes continuously. By now it is understood that out of equilibrium
these properties change in a rather dramatic manner. In general, when
the field $\rho({\bf x})$ is conserved in the bulk of the system,
$\phi\left[\rho({\bf x})\right]$ becomes a non-local functional \cite{spohn1983long,derrida2002large}.
This is directly related to the long-range correlations which are
known to exist in such systems for many years \cite{spohn1983long,dorfman1994generic,de2006hydrodynamic,bunin2013transport}.
Moreover, more recently it was realized that the functional can be
non-differentiable \cite{bertini2010lagrangian,bunin2012non,bunin2013cusp}.
Namely, for given configurations derivatives of $\phi\left[\rho({\bf x})\right]$
change discontinuously as $\rho(x)$ is varied. Such singularities
are well understood for many years in the context of low dimensional
systems \cite{graham1984existence,graham1984weak,graham1985weak,graham1986nonequilibrium,jauslin1987nondifferentiable}
and it is of interest to see 1) when they appear in continuum infinite
dimensional systems and 2) how they can be characterized. By now it
has been shown that the singularities occur for the weakly driven
asymmetric simple exclusion process (WASEP) in the limit of large
bulk driving field \cite{bertini2010lagrangian} and for a class of
boundary driven diffusive systems \cite{bunin2012non,bunin2013cusp}.
In the latter case, the singularities that were uncovered so far are
of a rather simple form that can be described by a mean-field Ising
like singularity (or in terms of catastrophe theory as a cusp singularity).
It was also shown that in the limit of small bulk bias the functional
becomes smooth as expected from the known results for the symmetric
simple exclusion process (SSEP). However, it is unclear which singularities
appear or how they can be described.

In this paper we take a closer look at singularities in the LDF of
the WASEP. As mentioned above the existence of singularities was first
shown in \cite{bertini2010lagrangian}. Specifically, the work proved
the existence of first order like singularities in the limit of a
large bulk driving field. However, the structure of the singularities
was not studied. Here, building on the results of \cite{bertini2010lagrangian},
we extend the study of singularities in the WASEP significantly using
a combination of numerical and analytical results. We first show,
numerically, how the singularities appear as the bulk bias is increased.
It is shown that for small bulk bias simple cusp like (or mean-field
Ising like) singularities appear. However, as the bulk bias increases
the structure of the singularities becomes more complicated and we
identify numerically an analogue of a tricritical point, also known
as a symmetry-restricted butterfly catastrophe \cite{gilmore1992catastrophe}.
For the cusp singularity we give evidence that it can be described
using a simple Landau like theory with an Ising symmetry, along the
lines of \cite{bunin2013cusp}. For the tricritical point the numerics
are not precise enough to verify that it can be described by a Landau
theory. We then consider the limit of an infinite bulk drive, namely,
the partially asymmetric simple exclusion process (PASEP) in the hydrodynamic
limit \cite{blythe2007nonequilibrium} (For an exact definition see
the discussion below). We characterize the singularity found in \cite{bertini2010lagrangian}
to show that the first order like singularity found there is a consequence
of a cusp singularity and extend the result to further show that analogues
of multicritical points of \emph{any order} appear. Finally we show
that in these can all be described using a Landau like theory.

The structure of the paper is as follows. In section \ref{sec:The-Model}
we describe the WASEP and PASEP models in the hydrodynamic limit.
In section \ref{Macro} we describe the Macroscopic Fluctuation Theory
and the solution obtained for the WASEP in \cite{bertini2010lagrangian}.
The numerical results for the singularities in the WASEP in the weak
bulk bias case are described in section \ref{sec:small-E}. In section
\ref{sec:infinite-E} we present analytical results for the PASEP,
including a mapping of the LDF of specific configurations to Landau
theory of multicritical points.

\section{\label{sec:The-Model}The Model}

\begin{figure}
\begin{centering}
\includegraphics[width=0.5\textwidth]{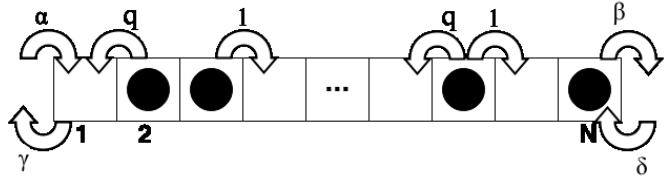} 
\par\end{centering}

\caption{\label{fig:microscopic-model} A pictorial illustration of the WASEP
and PASEP. Particles hope to the right (left) with rate $1$ ($q$),
provided that the site is empty. At the left (right) boundary particles
are injected with rate $\alpha$ ($\delta$) and removed with rate
$\gamma$ ($\beta$).}
\end{figure}

We consider the WASEP on a one-dimensional lattice with $N$ sites
(see Fig. \ref{fig:microscopic-model}). Particles hop to the left
with rate $q$ and to the right with rate $1$ (in arbitrary time
units), as long as the site that they hop to is not occupied. At the
left boundary, a particle is injected to site 1 with rate $\alpha$
(as long as site 1 is vacant), and if there is a particle at site
1, it is removed with rate $\gamma$. Similarly, at the right boundary,
a particle is injected to site $N$ with rate $\delta$, and removed
with rate $\beta$. When $q=1$, the bulk diffusion is unbiased, and
the process reduces to a SSEP.

The WASEP is defined in the limit when $1-q\sim N^{-1}$, meaning
that the bias strength scales inversely with system size. For this
case, in the hydrodynamic limit, the equation of motion for the particle
density field is \cite{spohn1991large} 
\begin{equation}
\partial_{t}\rho\left(x,t\right)+\partial_{x}J\left(x,t\right)=0,\label{eq:continuity}
\end{equation}
with 
\begin{equation}
J\left(x,t\right)=-\frac{1}{2}\partial_{x}\rho\left(x,t\right)+\sigma\left(\rho\right)E+\sqrt{\sigma\left(\rho\right)}\eta\left(x,t\right).\label{eq:current_dynamics}
\end{equation}
Here the spatial coordinate is rescaled by $N^{-1}$ and time by $N^{-2}$,
the diffusion coefficient has been set to $\nicefrac{1}{2}$ (in appropriate
units), $E$ is the bulk drive (proportional to $1-q$), and $\eta\left(x,t\right)$
is an uncorrelated white noise which satisfies $\left\langle \eta\left(x,t\right)\right\rangle =0$
and $\left\langle \eta\left(x,t\right)\eta\left(x',t'\right)\right\rangle =N^{-1}\delta\left(t-t'\right)\delta\left(x-x'\right)$,
with $N$ the system size. The dependence of the noise on $N$ is
a direct result of the rescaling of distances in the system. The noise
amplitude is given by $\sigma\left(\rho\right)=\rho\left(1-\rho\right)$
so that locally the equations of motion satisfy the fluctuation-dissipation
relation $\sigma(\rho)=k_{B}T\rho^{2}\kappa(\rho)$, where $\kappa(\rho)$
is the compressibility of a gas of diffusing hardcore particles, $T$
is the temperature and $k_{B}$ is the Boltzmann constant. The system
is attached to two reservoirs at $x=0$ and $x=1$ which impose the
boundary conditions 
\[
\rho(x=0)=\rho_{0}=\frac{\alpha}{\alpha+\gamma}\quad;\quad\rho(x=1)=\rho_{1}=\frac{\delta}{\delta+\beta}\ .
\]

\begin{figure}
\begin{centering}
\includegraphics[width=0.5\textwidth]{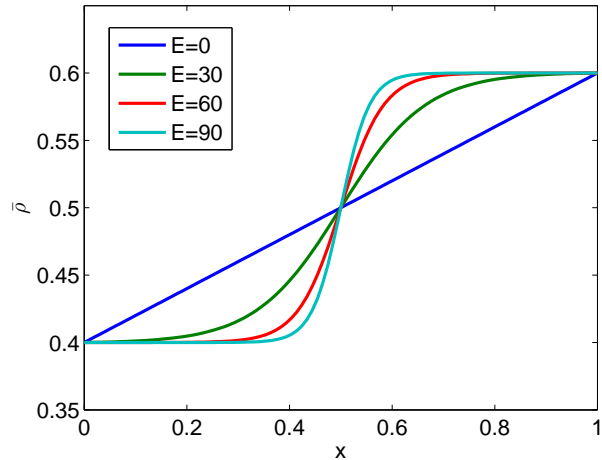} 
\par\end{centering}

\caption{\label{fig:Steady-states}Steady states of the WASEP for different
values of $E$. The boundary conditions are $\rho_{0}=0.4,\,\rho_{1}=0.6$.
As $E$ increases, the slope of the density profile in the middle
of the interval becomes steeper.}
\end{figure}

Throughout the paper our interest is in the case $\rho_{0}<\rho_{1}$
and $E>0$. Namely, the boundary conditions promote a particle current
in the negative $x$ direction and the field $E$ promotes a particle
current in the positive $x$ direction. The average density profile
$\bar{\rho}\left(x\right)$, which is also the most probable one,
is obtained by solving $-\frac{1}{2}\partial_{x}^{2}\bar{\rho}+\partial_{x}\sigma\left(\bar{\rho}\right)E=0$
with the boundary conditions $\rho_{0}$ and $\rho_{1}$. Fig. \ref{fig:Steady-states}
shows that as $E$ increases it changes from a linear density profile
($E=0$) to a step-like structure whose width scales, by dimensional
analysis, as $1/E$ (with the diffusion coefficient set to be $\nicefrac{1}{2}$).
In the limit $E\to\infty$ one reproduces the steady-state obtained
for the PASEP using a matrix product ansatz \cite{sasamoto1999one,blythe2000exact}.
Note that in the general the system is out of equilibrium, except
for the specific choice $E=\log\frac{\rho_{1}}{1-\rho_{1}}-\log\frac{\rho_{0}}{1-\rho_{0}}$,
for which that system is in equilibrium. In this case the average
current is equal to zero throughout the system.

To study the probabilities of large deviations for such a system we
use the macroscopic fluctuation theory (MFT) \cite{bertini2001fluctuations,bertini2002macroscopic}.
It will also be important for describing the structure and occurrence
of singularities in the LDF. To this end, in the next section we outline
the MFT for the WASEP, building on \cite{bertini2010lagrangian}.

\section{\label{Macro}Macroscopic Fluctuation Theory}

For our purpose it is most convenient to use a Hamiltonian approach.
To this end, we use a standard Martin-Siggia-Rose formalism \cite{martin_siggia_rose1973}.
Since we are interested in the steady-state probability density we
evaluate the probability of observing a certain density profile $\rho_{f}$
at time $t=0$, given that the system was at $\bar{\rho}$ at $t\rightarrow-\infty$.
This is given by 
\begin{equation}
P\left[\rho_{f}\left(x\right)\right]\sim\int\mathcal{D}\rho\mathcal{D}\eta\,\delta\left(\partial_{t}\rho+\partial_{x}J\right)\exp\left[-N\int_{-\infty}^{0}\mathrm{d}\tau\int_{0}^{1}\mathrm{d}x\frac{\eta^{2}}{2}\right],
\end{equation}
with the boundary conditions $\rho(x=0,t)=\rho_{0}$, $\rho(x=1,t)=\rho_{1}$,
$\rho(x,t=-\infty)=\overline{\rho}(x)$ and $\rho(x,t=0)=\rho_{f}(x)$.
Then following the standard procedure we introduce an auxiliary field
$\hat{\rho}$ and integrate over the noise to obtain 
\begin{equation}
P\left[\rho_{f}\left(x\right)\right]\sim\int\mathcal{D}\rho\mathcal{D}\hat{\rho}\,\exp\left[-N\int_{-\infty}^{0}\mathrm{d}\tau\int_{0}^{1}\mathrm{d}x\left(\hat{\rho}\partial_{t}\rho+\frac{1}{2}\partial_{x}\hat{\rho}\partial_{x}\rho-\partial_{x}\hat{\rho}\sigma\left(\rho\right)E-\frac{1}{2}\left(\partial_{x}\hat{\rho}\right)^{2}\sigma\left(\rho\right)\right)\right].\label{eq:MSR}
\end{equation}
As a result of boundary densities being fixed the field $\hat{\rho}$
satisfies the boundary conditions $\hat{\rho}(x=0)=0$ and $\hat{\rho}(x=1)=0$
\cite{tailleur2008mapping}.

In the large $N$ limit, which is of interest in this work, one evaluates
the path-integral using a saddle-point approximation. This yields
histories from the most probable configuration $\bar{\rho}\left(x\right)$
at $t=-\infty$ to $\rho_{f}(x)$ at $t=0$ which satisfy the Hamilton
equations 
\begin{eqnarray}
\partial_{t}\rho+\partial_{x}\sigma\left(\rho\right)E & = & \frac{1}{2}\partial_{x}^{2}\rho-2\partial_{x}\left(\sigma\left(\rho\right)\partial_{x}\hat{\rho}\right)\\
\partial_{t}\hat{\rho}+E\partial_{x}\sigma\left(\rho\right)\partial_{x}\hat{\rho} & = & -\left(\partial_{x}\hat{\rho}\right)^{2}\cdot\partial_{x}\sigma\left(\rho\right)-\frac{1}{2}\partial_{x}^{2}\hat{\rho}\;.
\end{eqnarray}
The LDF is then given by 
\begin{equation}
\phi[\rho_{f}(x)]=\inf_{i}\phi_{i}[\rho_{f}(x)]\label{eq:LDF_infimum}
\end{equation}
where 
\begin{equation}
\phi_{i}[\rho_{f}(x)]=\phi[\rho_{f}(x),\rho_{i}\left(x,t\right)],\label{eq:LDF}
\end{equation}
and 
\begin{eqnarray}
\phi[\rho_{f}(x),\rho_{i}\left(x,t\right)] & \equiv & \int_{-\infty}^{0}\mathrm{d}\tau\int_{0}^{1}\mathrm{d}x\left(\hat{\rho_{i}}\partial_{t}\rho_{i}+\frac{1}{2}\partial_{x}\hat{\rho_{i}}\partial_{x}\rho_{i}-\partial_{x}\hat{\rho_{i}}\sigma\left(\rho_{i}\right)E-\frac{1}{2}\left(\partial_{x}\hat{\rho_{i}}\right)^{2}\sigma\left(\rho_{i}\right)\right)\label{eq:LDF_action}
\end{eqnarray}
is an action analogue evaluated at the saddle-point solution $h_{i}\equiv(\rho_{i}(x,t),\,\hat{\rho_{i}}(x,t))$.
Note that we have allowed in our notations for multiple saddle point
solutions, labeled by $i$. Since we are looking for the infimum (Eq.
\ref{eq:LDF_infimum}), we will consider only solutions that are local
minima. As we will see, when more than one minimizing history exists
the LDF can exhibit singularities \cite{bertini2010lagrangian,bunin2012non}.

As noted in \cite{bertini2010lagrangian} in the case of the WASEP
it is useful to perform a canonical transformation to the variables:
\begin{eqnarray}
\varphi\left(x\right) & = & \log\left(\frac{\rho(x)}{1-\rho(x)}\right)-\hat{\rho}\left(x\right)\;,\label{eq:varphi}\\
\psi\left(x\right) & = & \rho\left(x\right)\ ,
\end{eqnarray}
with $0\leq x\leq1$ and which satisfy the equations: 
\begin{eqnarray}
\varphi_{t} & = & \varphi_{xx}-\left(1-2\psi\right)\varphi_{x}\left(E-\varphi_{x}\right)\ ,\\
\psi_{t} & = & -\psi_{xx}-E\left[\psi\left(1-\psi\right)\right]_{x}+2\left[\psi\left(1-\psi\right)\varphi_{x}\right]_{x}\ .
\end{eqnarray}
Solving for $\psi$ one obtains 
\begin{equation}
\rho=\psi=\frac{1}{1+e^{\varphi}}-\frac{\varphi_{xx}}{\varphi_{x}\left(E-\varphi_{x}\right)}\label{eq:phi_to_rho_mapping}
\end{equation}
which can be used to obtain a single equation for the time evolution
of the field $\varphi$ 
\begin{equation}
\varphi_{t}=-\varphi_{xx}+\frac{1-e^{\varphi}}{1+e^{\varphi}}\varphi_{x}\left(E-\varphi_{x}\right).\label{eq:phi_dynamics}
\end{equation}
The boundary conditions on $\varphi(x)$ are $\varphi\left(x=0\right)=\log\frac{\rho\left(x=0\right)}{1-\rho\left(x=0\right)}$
and $\varphi\left(x=1\right)=\log\frac{\rho\left(x=1\right)}{1-\rho\left(x=1\right)}$
(see Eq. \ref{eq:varphi} and recall that $\hat{\rho}$ vanishes at
the boundaries).

Using these results an exact expression for the LDF in the infinite
$E$ limit was recovered in \cite{bertini2010lagrangian} (see Sec.
\ref{sec:infinite-E}). As expected, the result agrees with the expression
obtained using other methods \cite{derrida2003exact,enaud2004large}.
However, the structure of the resulting LDF has not been explored
in detail. An exception are the results of Bertini et. al. \cite{bertini2010lagrangian}
which showed that in the large $E$ limit, for a range of configurations,
the saddle-point solutions can have three solutions, two locally stable
and one locally unstable. Generically, one of the locally minimizing
solutions gives a lower LDF value than the other and therefore controls
the probability distribution. However, as the configuration $\rho_{f}$
is changed there are certain values of $\rho_{f}$ for which the two
locally minimizing histories give the same value of the action. At
these points, much like a first order phase transition, the history
which controls the probability distribution changes and the LDF become
singular. In the limit of $E=0$ it is well known \cite{derrida2007non}
that the saddle-point equations admit only one solution. Indeed, as
noted by Bertini et. al. for small enough values of $E$ these singularities
disappear.

In what follows we build on the results obtained by Bertini et. al.
and explore the structure of the LDF in much more detail. We show
that rather complicated singular structures, with \textit{more} than
two stable solutions, can also occur and characterize their structure.
Importantly, we show that these can be analyzed using a Landau like
theory. Furthermore, we show, for small values of the field, how the
different singular structures emerge as the magnitude of $E$ is increased
(and the length scale $E^{-1}$ decreases). To this end, in what follows
we will use the results presented above numerically in the small $E$
limit and then analytically in the infinite $E$ limit.

\section{\label{sec:small-E}The small $E$ limit}

To study the small $E$ limit we first use Eq. \ref{eq:phi_to_rho_mapping}
to express the field $\rho_{f}\left(x\right)$ in terms of $\varphi\left(x,t=0\right)$,
and $\bar{\rho}(x)$ in terms of $\varphi\left(x,t=-\infty\right)$.
We then solve \textbf{numerically} for the dynamics of $\varphi\left(x,t\right)$
using Eq. \ref{eq:phi_dynamics} and use the result to obtain the
history $\rho(x,t)$ using Eq. \ref{eq:phi_to_rho_mapping}. The result
allows us to evaluate the LDF for specific values of $\rho_{f}(x)$.
The main advantage of this procedure is the relative ease in which
one can identify cases when multiple saddle-point solutions exist
(as stated above these can lead to LDF singularities which are the
subject of this paper). Specifically, the mapping, Eq. \ref{eq:phi_to_rho_mapping},
being a nonlinear boundary value problem, may have multiple solutions
of $\varphi$ for the same value of $\rho_{f}$ \cite{bernfeld1974introduction}.
Each such solution generates a distinct extremal history (which can
be either a local minimum or a local maximum). Since the equation
of motion for $\varphi$ is an initial value differential equation
it cannot have multiple solutions on its own. Therefore, using this
procedure reduces the problem of finding multiple histories of the
Hamilton time dependent equations to finding multiple solutions of
a time independent differential equation.

\begin{figure}
\includegraphics[width=0.5\textwidth]{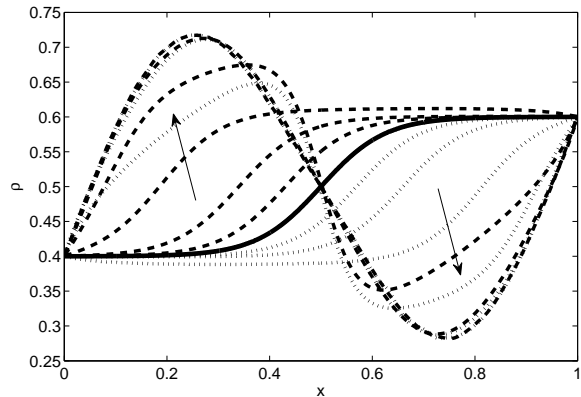}

\caption{\label{fig:trajectories_E40}Two histories starting at the steady
state profile (solid line), and leading to the same density profile
with $a_{2}=0.3$, $a_{1}=0$, $a_{4}=0$ in Eq. \ref{eq:finite_profile_3}
and $E=40$, $\rho_{0}=0.4$ with $\rho_{1}=0.6$. The two histories
have the same statistical weight. One of the histories is depicted
by dashed lines, and the other one by a dotted lines. Each line represents
the density profile at a specific time, with the arrows showing the
direction of the evolution in time.}
\end{figure}

Scanning of the full configuration space of the field $\rho_{f}(x)$
is impossible. Therefore, similar to \cite{bunin2012non}, we constrain
ourselves to finite-dimensional cuts. In particular, we first focus
on smooth long wave length structures. To this end, we consider configurations
of the form 
\begin{equation}
\rho_{f}\left(x\right)=\bar{\rho}(x)+a_{1}\sin\left(\pi x\right)+a_{2}\sin\left(2\pi x\right)+a_{4}\sin\left(4\pi x\right)\ .\label{eq:finite_profile_3}
\end{equation}
$a_{1}$, $a_{2}$ and $a_{4}$ loosely measure the size of the deviation
from the most probable configuration $\bar{\rho}(x)$ at different
wavelengths. Their values are constraint since the density is bound
between $0\leq\rho(x)\leq1$. We have verified that the exact choice
of the functions (sine or other) describing the long wavelength behavior
is not important for the overall structure of the results presented.

Our interest, as stated above, is identifying configurations at which
the LDF is singular. As will become evident, to do so it is useful
to employ the symmetries of the problem. We consider boundary conditions
such that 
\begin{equation}
\rho_{0}=0.5-\delta\quad;\quad\rho_{1}=0.5+\delta\ .\label{eq:symmetric_boundary_conditions}
\end{equation}
Note that under this choice of boundary conditions $\bar{\rho}(x)$
satisfies a particle-hole symmetry so that under the exchange $x\to-x$
and $\rho\to1-\rho$ the most probable profile returns to itself.
While a structure similar to what we find emerges for other choices
of boundary conditions the results have a simpler form for this choice.
In particular, for this choice $a_{2}$ and $a_{4}$ are deviations
from $\bar{\rho}(x)$ which satisfy the particle-hole symmetry while
$a_{1}$ breaks the symmetry.

Following the procedure outlined above we use the mapping, Eq. \ref{eq:phi_to_rho_mapping},
to scan systematically, on the finite dimensional cuts, for cases
where multiple saddle point solutions occur by looking for multiple
$\varphi$ solutions of the differential equation for the same $\rho_{f}$.
This is carried out numerically by using an extended `shooting' algorithm
\cite{press2007numerical} whose details are given in \ref{sec:appendix-shooting}.
For the purpose of the discussion here we note that eventually the
solutions are obtained by breaking the interval $\left[0,1\right]$
to $L$ bins. The accuracy of the solution increases with $L$.

We now describe the singular structures which we identify using this
method.

\subsection{The appearance and characterization of Ising-like (cusp) singularities}

\begin{figure}
\includegraphics[width=0.5\textwidth]{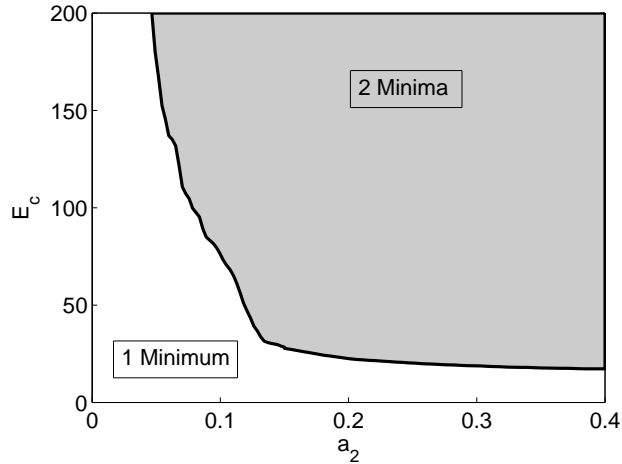}

\caption{\label{fig:cusp_disappearence}Minimum field $E_{c}$ required for
an Ising-like (cusp) singularity, versus amplitude of the sine for
the profile $\rho_{f}\left(x\right)=\bar{\rho}\left(x\right)+a_{2}\cdot\sin\left(2\pi x\right)$,
where $\bar{\rho}\left(x\right)$ is the average profile for the PASEP
with $\rho_{0}=0.4,\,\rho_{1}=0.6$. In the shaded area, two degenerate
minimizing histories coexist. The border line is a second-order transition
line.}
\end{figure}

Consider first profiles such that $a_{1}=0$ and $a_{4}=0$ with $a_{2}$
non-zero. Such profiles are particle-hole symmetric. We find numerically
that for large enough values of $E$ there is a critical value of
$a_{2}$ for which multiple solutions of the saddle-point equations
appear. An example is shown in Fig. \ref{fig:trajectories_E40} where
the two locally minimizing histories are shown (an extra locally maximizing
solution is also present but not shown). Note that, due to the symmetry,
the two histories are connected through a particle-hole symmetry transformation,
and both give the same value for the action in Eq. \ref{eq:LDF_action}.
In Fig. \ref{fig:cusp_disappearence} we show the minimal value of
$E$, denoted by $E_{c}$, for which two minimizing solutions appear
for different values of $a_{2}$ for a specific choice of $\rho_{0}$
and $\rho_{1}$ (we have verified that the qualitative results are
insensitive to this choice). Note that 1) there is a minimal value
of $a_{2}$ above which degenerate, namely with equal values of the
action, solutions appear and 2) there is a minimal value of the field
$E$ below which a singularity never appears in the LDF (this value
is considerably higher than the value of the field, $E\simeq0.81$
for this choice of boundary conditions, at which the system is in
equilibrium). Namely, singularities of the LDF appear only for conditions
where the field $E$ is large enough, and for ``large enough'' deviations
from the most probable configuration.

\begin{figure}
\includegraphics[width=0.5\textwidth]{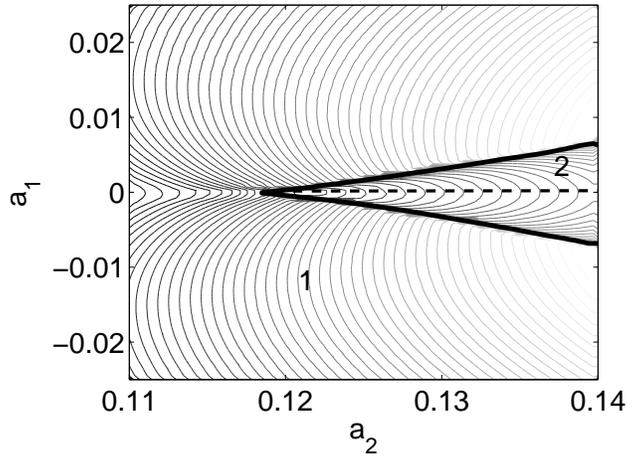}

\caption{\label{fig:finite_e_cusp}An Ising-like (cusp) singularity in the
small E regime. Along the plotted lines the value of LDF is constant.
The field strength is $E=40$ and the boundary conditions are $\rho_{l}=0.4\,;\,\rho_{r}=0.6$.
The solid line is the border between the area with one minimum and
the area with two minima. The dashed line is a first order transition
line. On that line, two distinct histories, which have the same statistical
weight, minimize the action. All the lines meet at an analogue of
a critical point (cusp).}
\end{figure}

The plane depicted in Fig. \ref{fig:cusp_disappearence} contains
a region in configuration space where two degenerate histories, which
we denote by $h_{1}$ and $h_{2}$ (in the sense that they lead to
the same value of the action from Eq. \ref{eq:LDF_action}) coexist.
If we make $a_{1}>0$ but small, the particle-hole symmetry is broken
between the two histories and now one of the histories, say $h_{1}$,
leads to a lower value of the action than the other. On the other
hand for $a_{1}<0$, $h_{2}$ leads to a lower value of the action.
Therefore, at $a_{1}=0$ there is a first-order-line singularity of
the LDF. For a given value of $E$, as $a_{2}$ is decreased the two
distinct histories merge into a single history, much like a critical
point in usual phase transitions (or a cusp catastrophe). These results
are illustrated in Fig. \ref{fig:finite_e_cusp} where the resulting
large deviation for a given value of $E$ in the $a_{1}$, $a_{2}$
plane is shown. Note that as $a_{1}$ is increased (for large enough
$a_{2}$) the two solutions initially coexist (with one dominating
the LDF) until eventually for large enough $a_{1}$ one of the solutions
disappears. We comment that because of numerical precision seeing
the singularity in the plotted lines of equal LDF value is rather
hard. Their existence is most easily obtained by tracking where solutions
appear and disappear.

To describe the singularities we follow \cite{bunin2013cusp} and
use a Landau like theory with a $Z_{2}$ symmetry. We look at the
behavior of the LDF, $\phi\left[\rho\right]$, in the vicinity of
the critical configuration $\rho_{f}^{cusp}$. As stated before, for
a given $\rho_{f}$ on the switching line, there are two minimizing
degenerate histories, $h_{1}\left(x,t\right)$ and $h_{2}\left(x,t\right)$.
To build the Landau theory we introduce 
\begin{equation}
a=\left[\int\left(\rho_{f}-\rho_{f}^{cusp}\right)^{2}\mathrm{d}x\right]^{\nicefrac{1}{2}}\label{eq:order_param_a}
\end{equation}
as the distance of the configuration $\rho_{f}$ from $\rho_{f}^{cusp}$.
Then we define a coordinate system $\left(a,b\right)$ with $\rho_{f}^{cusp}$
at the origin, $\hat{a}$ directed along the switching line and positive
on the switching line, and $\hat{b}$ orthogonal to $\hat{a}$. In
analogy with Landau mean-field theory, $a$ plays the role of the
temperature `distance' from the critical point and $b$ the role of
the magnetic field. Let 
\begin{eqnarray}
\rho^{avg}\left(x,t\right) & = & \frac{1}{2}\left[h_{1}\left(x,t\right)+h_{2}\left(x,t\right)\right],\nonumber \\
\delta\rho\left(x,t\right) & = & \frac{1}{2}\left[h_{1}\left(x,t\right)-h_{2}\left(x,t\right)\right],\\
u\left(x,t\right) & = & \delta\rho/\left\Vert \delta\rho\right\Vert \nonumber 
\end{eqnarray}
and 
\begin{equation}
\Delta=\left\Vert \delta\rho\right\Vert \label{eq:order_param_cusp}
\end{equation}
where $\left\Vert \delta\rho\right\Vert ^{2}=\int\left[\delta\rho\left(x,t\right)\right]^{2}\mathrm{d}x\mathrm{d}t$
quantifies the distance between the two histories. Note that at the
cusp, where the two histories coincide, $\Delta=0$. As will shortly
become clear $\Delta$, which measures the distance between the two
histories, is the order parameter of the Landau theory.

On the switching line $b=0$ and $\phi\left[\rho_{f},h_{1}\right]$
and $\phi\left[\rho_{f},h_{2}\right]$ are both minimizing histories
with the same weight. Hence the function 
\begin{equation}
s_{\rho_{f}}\left(q\right)=\phi\left[\rho_{f},\,\rho^{avg}+q\cdot u\right]
\end{equation}
admits two minima, at $q=\pm\Delta$. In order to capture this behavior
of two minima converging to one at a `critical' point, we use the
simplest analytical form possible: 
\begin{equation}
\tilde{s}\left(q\right)=s\left(q\right)-s\left(0\right)=c_{4}q^{4}+ac_{2}q^{2}+bc_{1}q,
\end{equation}
with $c_{1},c_{2},c_{4}>0$. $q$ is a selector between histories.
Being so, it contains information which is inherently non-local both
in time and space. At small $a$ and $b=0$, $\tilde{s}\left(q\right)$
has two minima, at $q_{min}\propto\pm\sqrt{a}$. Hence $\Delta\propto\sqrt{a}$
in direct analogy with a Landau theory, with an order parameter critical
exponent $\beta$ with a value of $\nicefrac{1}{2}$.

\begin{figure}
\includegraphics[width=0.5\textwidth]{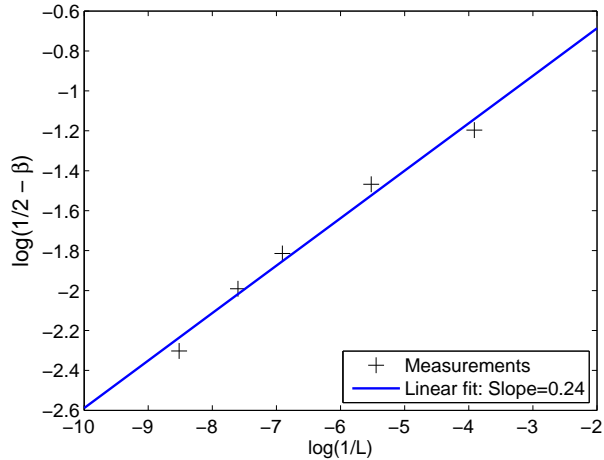}

\caption{\label{fig:cusp-critical-exponent}Critical exponent measurements
for the cusp singularity in the WASEP model. The graph shows the approach
of the critical exponent $\beta$, for different numerical precisions.
$L$ is the number of bins taken inside the interval $\left[0,1\right]$.
See main text for more details. One can witness that as the numerical
precision rises the critical exponent approaches a value of 0.5.}
\end{figure}

In order to test the analogy, we generated a log-log plot of $\Delta\left(a\right)$,
and measured the slope, $\beta$. In Fig. \ref{fig:cusp-critical-exponent}
we plot the value of the resulting measured exponent $\beta$ as a
function of the number of bins, $L$, used in the numerics. The results
strongly suggest that $\beta\rightarrow\nicefrac{1}{2}$ as $L\rightarrow\infty$.

\subsection{Tricritical like singularities}

\begin{figure}
\includegraphics[width=0.5\textwidth]{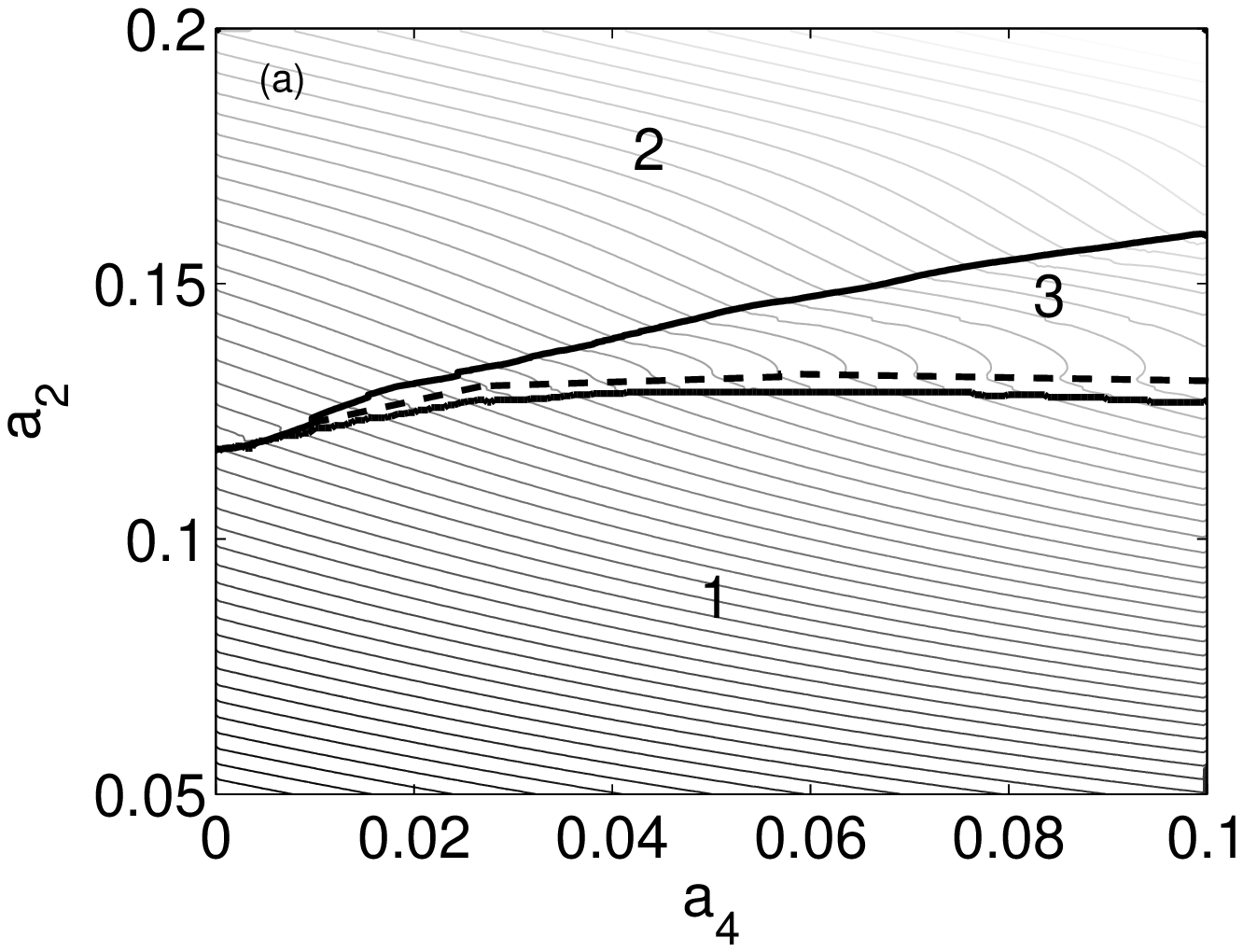}

\includegraphics[width=0.5\textwidth]{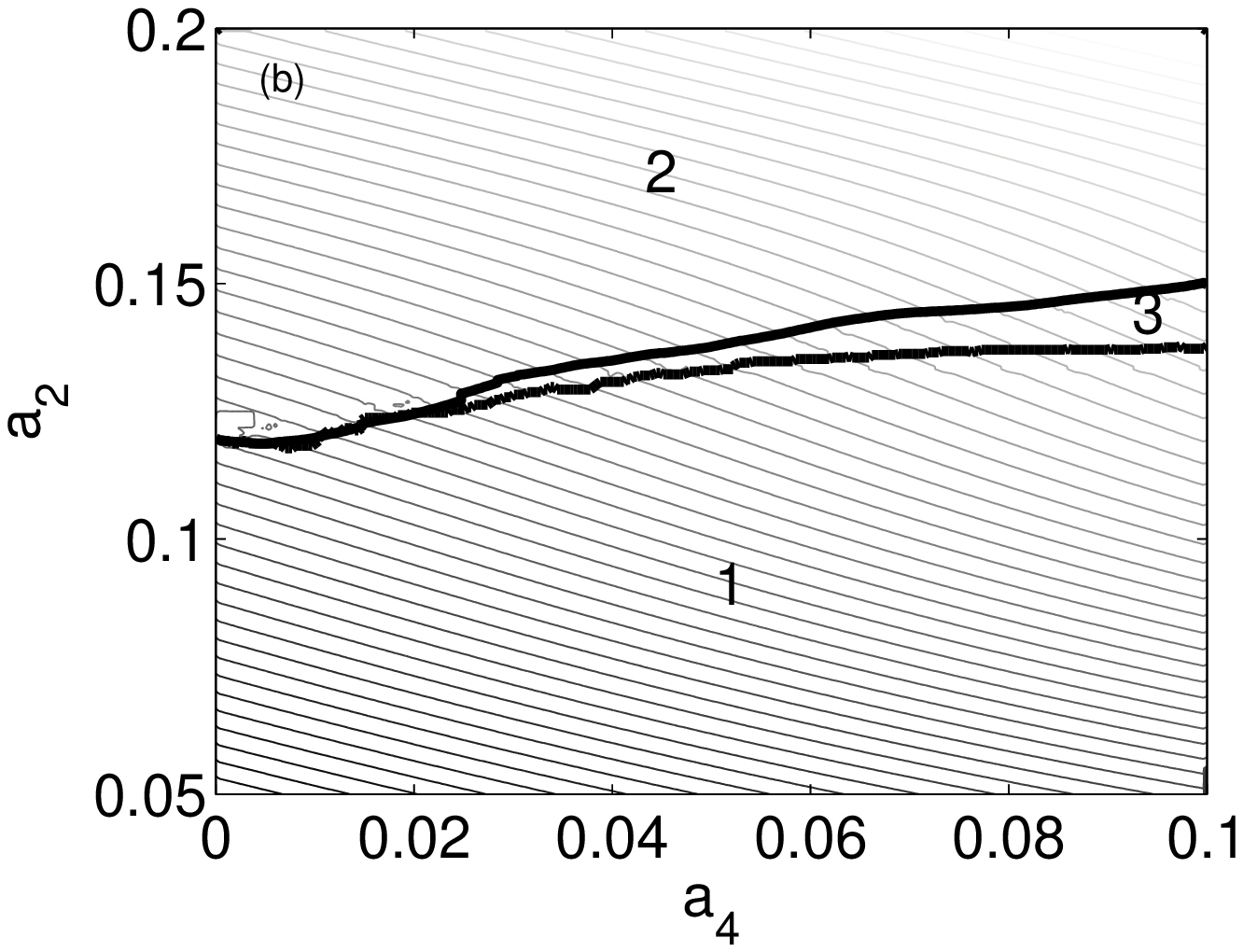}

\includegraphics[width=0.5\textwidth]{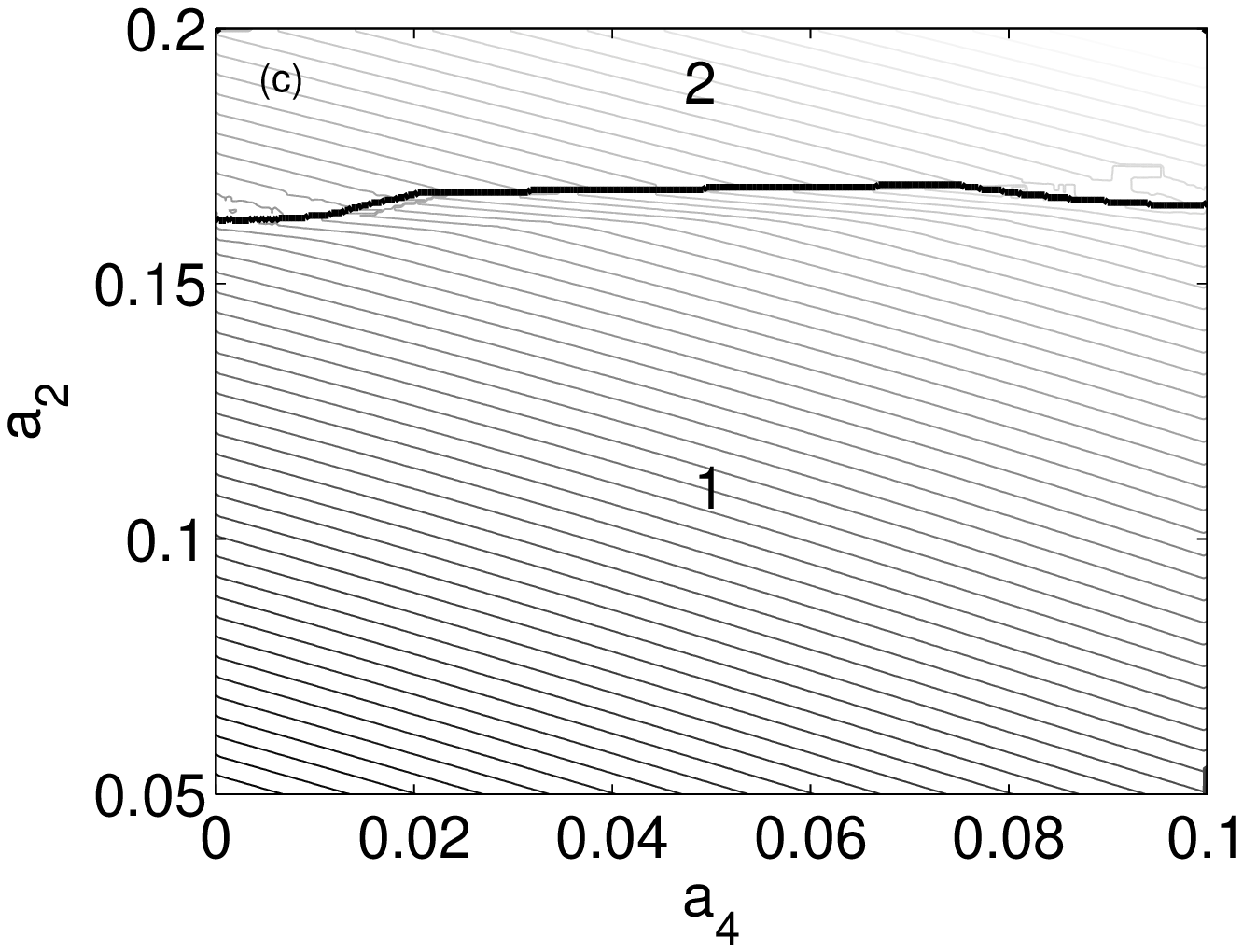}

\caption{\label{fig:PASEP_tricritical_decay}`Phase diagrams' in the $a_{2},\, a_{4}$
plane for (a) $E=50$, (b) $E=40$ and (c) $E=25$ (from top to bottom).
The numbers denote the number of locally minimizing solutions in the
plane. Along the thin line the value of the LDF is constant. The thick
solid lines are borders between areas with different numbers of locally
minimizing histories (they are not necessarily transition lines).
The tip of the three minima area is where the tricritical-like point
resides. In (a) the dashed line (a guide to the eye) is a first order
transition line where all three minima give the same value for the
action, and the LDF shows a singularity structure manifested in a
clear ``break'' of the equipotential lines. Below this line, the
single minimum has the lowest value, and above this line the two degenerate
minima have a lower value and they determine the value of the LDF.
The border between the one minimum and two minima area is a second
order transition line. In (b) the first order transition line almost
overlaps the line separating the one and the three minima areas, so
we do not show it. Note how the three minima area is not seen for
(c) $E=25$ and only a cusp singularity remains (the transition is
on the boundary between one and two minima).}
\end{figure}

It is natural to ask if there exist more complicated situations, where
more than two solutions coexist. Indeed, as we now show, as the strength
of the field is increased we find that a region with five extremal
coexisting solutions (three locally minimizing and two locally maximizing)
appears. In Fig. \ref{fig:PASEP_tricritical_decay} we show for different
values of $E$ the number of solutions in the $a_{2}$, $a_{4}$ plane
for $a_{1}=0$. Namely, final configurations with a particle-hole
symmetry. Note that due to the symmetry on this plane solutions related
by particle-hole symmetry are degenerate.

For small $E$ we see that there is only one solution for each final
configuration (data not shown). As $E$ increases a region with two
locally minimizing solutions appears as described in the previous
section. More interesting, for larger $E$ we find a region with three
locally minimizing solutions (of a total of five solutions). For the
smaller values of $E$ it is present only for large enough values
of $a_{2}$ and $a_{4}$ (large enough deviations) while as $E$ increases
the region covers a larger portion of the two dimensional cut. When
three locally minimizing solutions are present we find that one obeys
the particle-hole symmetry and the other two break it (data not shown).
The structure is very close to that which emerges from a Landau theory
of a tricritical point. The tricritical point occurs when the regions
with one, two and three solutions meet.

Note that inside the three solutions area a line where the LDF is
singular appears. This line is a transition due to a competition between
the one minima with a particle-hole symmetric time evolution and the
other two which are degenerate and break the particle hole symmetry.
This behavior is, again, in direct analogy to a Landau theory of a
tricritical point. The transition between the two is first-order.
The transition line meets a second-order transition line as in a usual
tricritical point structure. Note that along the lines where the region
with three solutions turns into a two solution region the LDF is not
singular but changes smoothly. Again, as in the discussion of the
Ising-like singularity, inferring the singularities from the lines
of equal LDF value can be misleading and it is best to track the number
of solutions and their behavior. Therefore within the numerics we
can only estimate the location of the singularities. Indeed, for intermediate
values of the field even at regions where $3$ solutions were clearly
visible it was numerically hard to identify the first order line.

In direct analogy to the previous section also here a Landau theory
can be constructed. However, our numerics are not good enough to verify
the expected exponents associated with the tricritical point.

In the next section we show that when $E\to\infty$ tricritical point
type singularities (and much more complicated) appear. In that case
the structure of the singularity is described exactly by a Landau-like
theory.

\section{\label{sec:infinite-E}The LDF at infinite bulk drive}

Next, we turn to consider the limit $E\to\infty$. In the common terminology
this corresponds to a PASEP and as before $E$ is in the positive
direction with $\rho_{1}>\rho_{0}$. Similar to the above discussion
we consider specific cuts of the configuration space $\rho(x)$. The
main points of the discussion which follows are: 1) Within the subspaces
studied we can identify configurations at which the LDF is singular.
2) Around appropriate configurations the singular behavior of the
LDF can be described exactly by a Landau like theory. The simplest
ones will be, as above, cusp (or Ising) singularities and an analogue
of a tricritical point. More striking is the identification of configurations
at which an arbitrary number of histories reach the same final configuration
with the same weight. This leads to an analogue of an arbitrary order
multicritical point. We note that while an exact correspondence between
the order parameter defined in the previous section and the one used
in this section is not shown their general behavior is identical.

To show the above results we use the $E\to\infty$ limit expression
for the LDF obtained in \cite{derrida2003exact,bertini2010lagrangian}.
There it was shown that 
\begin{equation}
\phi\left[\rho_{f}\left(x\right)\right]=-\mathcal{K}\left(\rho_{0},\rho_{1}\right)+\mathcal{F}\left[\rho_{f}\left(x\right)\right]+\inf_{0<y<1}\mathcal{G}\left[\rho_{f}\left(x\right),y\right]\ ,\label{eq:PASEP_LDF}
\end{equation}
where 
\begin{eqnarray}
\mathcal{K}\left(\rho_{0},\rho_{1}\right) & = & \min\left[\log\rho_{0}\left(1-\rho_{0}\right),\log\rho_{1}\left(1-\rho_{1}\right)\right]\ ,\\
\mathcal{F}\left[\rho_{f}\left(x\right)\right] & = & \int_{0}^{1}\mathrm{d}x\left[\rho_{f}\left(x\right)\log\left(\rho_{f}\left(x\right)\right)+\left(1-\rho_{f}\left(x\right)\right)\log\left(1-\rho_{f}\left(x\right)\right)\right]\ ,\nonumber 
\end{eqnarray}
and $\mathcal{G}$ is given by 
\begin{eqnarray}
\mathcal{G}\left[\rho_{f}\left(x\right),y\right] & = & \int_{0}^{y}\mathrm{d}x\left[\rho_{f}\left(x\right)\log\left(1-\rho_{0}\right)+\left(1-\rho_{f}\left(x\right)\right)\log\left(\rho_{0}\right)\right]\label{eq:PASEP_QuasiLDF}\\
 & + & \int_{y}^{1}\mathrm{d}x\left[\rho_{f}\left(x\right)\log\left(1-\rho_{1}\right)+\left(1-\rho_{f}\left(x\right)\right)\log\left(\rho_{1}\right)\right]\ .\nonumber 
\end{eqnarray}
The result can be obtained by taking the $E\to\infty$ limit using
the results of Sec. \ref{Macro}.

Clearly, any singular behavior of the functional can appear only in
$\mathcal{G}$. To this end, it is convenient to only consider the
behavior of 
\begin{equation}
g\left[\rho_{f}\left(x\right)\right]=\inf_{y}\mathcal{G}\left[\rho_{f}\left(x\right),y\right]\,.
\end{equation}
The discussion of Sec. \ref{Macro} can be shown to imply that for
a given final configuration $\rho_{f}(x)$ each locally minimizing
value of $y$ corresponds to a particular choice of history which
leads to it \cite{bertini2010lagrangian}.

Next, for simplicity we consider density profiles of the form 
\begin{equation}
\rho_{f}\left(x\right)=\left(\frac{1}{2}-\delta\right)+2\delta\cdot x+\sum_{n=1}^{2n_{max}}a_{n}\sin\left(n\pi x\right)\ ,\label{eq:profiles}
\end{equation}
where $\delta\in\left(0,\nicefrac{1}{2}\right]$ and use the boundary
conditions of Eq. \ref{eq:symmetric_boundary_conditions}. It is straightforward
to show that profiles which cross $\rho=\nicefrac{1}{2}$ $k$ times
have $k$ extremal histories leading to them \cite{bertini2010lagrangian}.
Note, that $\rho_{f}$ has an implicit dependence on $n_{max}$ which
we suppress most of the time for brevity. These profiles will allow
us to look at different subspaces of $\phi\left[\rho_{f}\left(x\right)\right]$
by characterizing the function $\rho_{f}\left(x\right)$ using the
vector $\left(a_{1},...,a_{2n_{max}}\right)$. While singularities
are likely to occur for other configurations and boundary conditions,
this particular choice allows for a particle-hole symmetry to be exploited
and analytical results to be obtained. This gives a simple mapping
of the problem to a Landau theory around specific values of $\left(a_{1},...,a_{2n_{max}}\right)$.

\subsection{Mapping to Landau theory of different types}

We start by considering profiles where cusp and tricritical-like singularities
appear. As before, for cusp singularities there is a region in configuration
space where two locally minimizing histories lead to the same final
configuration. The Landau like expansion is carried around the point
where the two histories merge into one. For the tricritical point
there is a region where three locally minimizing solutions lead to
the same final configuration and the expansion is carried around the
point where the three histories merge into one. Finally, the generalization
to higher order cases will be derived.

To carry out the mapping we define $m=y-1/2$ and expand $\mathcal{G}(m)$
in powers of $m$. A straightforward calculation shows that 
\begin{equation}
\mathcal{G}\left(m\right)=2\log\left(\frac{1+\delta}{1-\delta}\right)\cdot\sum_{n=0}^{\infty}c_{n}m^{n}\,.\label{eq:G_expansion}
\end{equation}
Here $c_{0}$ is a constant (which can be ignored) and we have extracted
the constant $2\log\left(\frac{1+\delta}{1-\delta}\right)$ to simplify
the expressions for the coefficients $c_{n}$. Using Eq. \ref{eq:profiles},
it can be shown that \numparts 
\begin{eqnarray}
c_{1}= & \frac{1}{2}\left(2\left.\rho\right|_{m=0}-1\right)=\sum_{n\in odd}\left(-1\right)^{\frac{n-1}{2}}a_{n},\label{eq:multicritical_c1}\\
c_{2}= & \frac{1}{2}\left.\frac{\mathrm{d}}{\mathrm{d}m}\rho\right|_{m=0}=\frac{1}{2}\left[2\delta+\sum_{n\in even}\left(-1\right)^{\frac{n}{2}}n\pi a_{n}\right],\label{eq:multicritical_c2}
\end{eqnarray}
\endnumparts and more generally (for $k>1$) \numparts \label{eq:multicritical_c_all}
\begin{eqnarray}
c_{2k-1}= & \frac{1}{\left(2k-1\right)!}\left.\frac{\mathrm{d}^{2k-2}}{\mathrm{d}m^{2k-2}}\rho\right|_{m=0}=\frac{\left(-1\right)^{k-1}}{\left(2k-1\right)!}\sum_{n\in odd}\left(-1\right)^{\frac{n-1}{2}}\left(n\pi\right)^{2k-2}a_{n}\,.\label{eq:multicritical_c_odd}
\end{eqnarray}
\begin{eqnarray}
c_{2k}= & \frac{1}{\left(2k\right)!}\left.\frac{\mathrm{d}^{2k-1}}{\mathrm{d}m^{2k-1}}\rho\right|_{m=0}=\frac{\left(-1\right)^{k+1}}{\left(2k\right)!}\sum_{n\in even}\left(-1\right)^{\frac{n}{2}}\left(n\pi\right)^{2k-1}a_{n}\,,\label{eq:multicritical_c_even}
\end{eqnarray}
\endnumparts Notice that $c_{n}$ with $n$ odd include only $a_{n}$
coefficients with $n$ odd and similarly $c_{n}$ with $n$ even include
only $a_{n}$ with $n$ even. This is a direct consequence of the
choice of boundary conditions and profiles. Profiles which involve
only $a_{n}$ with even values of $n$ have a particle-hole symmetry
while those with $a_{n}$ with $n$ odd break this symmetry.

Finally, note that $g\left[\rho_{f}\left(x\right)\right]=\inf_{m}\mathcal{G}\left[\rho_{f}\left(x\right),\, m\right]$
in analogy with a Landau free energy with $m$ playing the role of
the order parameter. We now turn to discuss specific configurations
where a singular LDF appears.

\subsubsection{Ising-like or Cusp singularities:}

Here we consider profiles of the form: 
\begin{equation}
\rho_{f}\left(x\right)=\left(\frac{1}{2}-\delta\right)+2\delta\cdot x+a_{1}\sin\left(\pi x\right)+a_{2}\sin\left(2\pi x\right)\ .\label{eq:profile_cusp}
\end{equation}

Substituting this particular choice into Eqs. \ref{eq:multicritical_c_all}
we obtain \numparts 
\begin{eqnarray}
c_{1}= & a_{1}\,,\\
c_{2}= & \delta-\pi a_{2}\,,\\
c_{3}= & -\frac{\pi^{2}}{6}a_{1}\,,\\
c_{4}= & \frac{\pi^{3}}{3}a_{2}\,.
\end{eqnarray}
\endnumparts While in general $c_{n}$ with $n>4$ appear it is clear
that by choosing $c_{2}$ and $c_{1}$ small they can be neglected.
Using the standard arguments of the Landau theory the singular behavior
of the LDF can be captured by 
\begin{equation}
\mathcal{G}\left(m\right)\simeq2\log\left(\frac{1+\delta}{1-\delta}\right)\cdot\left(c_{0}+a_{1}m+(\delta-\pi a_{2})m^{2}+\frac{\pi^{3}}{3}a_{2}m^{4}\right)\;,\label{eq:G_expansion_cusp}
\end{equation}
where the expansion is taken about the critical-point 
\begin{equation}
\left(a_{1}^{\star},\, a_{2}^{\star}\right)=\left(0,\,\frac{\delta}{\pi}\right)\ .
\end{equation}
Note that the coefficient of $m^{4}$ is positive near the critical
point. Following a standard Landau theory it is clear that the structure
of $\mathcal{G}\left(m\right)$ implies that there is a first-order
like transition line (on which the derivative of the LDF has a discontinuity)
ending in a critical-point analogue. Furthermore, as before this implies
that approaching the critical point along this line the minimizing
value of $m$, denoted by $m^{\star}$ gives $m^{\star}\propto(\delta-\pi a_{2})^{\nicefrac{1}{2}}$
with other standard Landau theory results following.

\begin{figure}[h]
\includegraphics[width=0.5\textwidth]{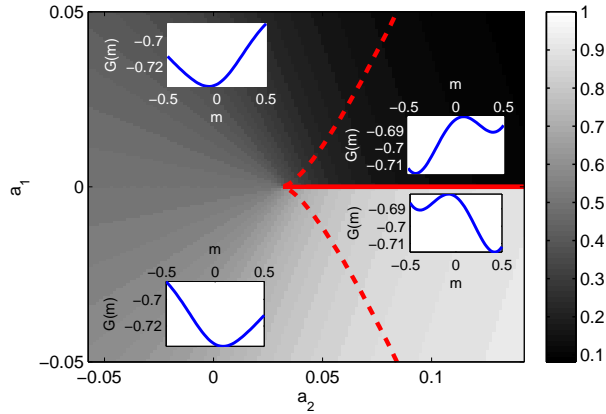}

\caption{\label{fig:full_cusp} The `phase diagram' for profiles of the form
$\rho_{f}\left(x\right)=\frac{1}{2}-\delta+2\delta\cdot x+a_{1}\sin\left(\pi x\right)+a_{2}\sin\left(2\pi x\right)$
with $\delta=0.1$, which demonstrates an Ising singularity (or cusp
catastrophe). The background color represents the value of the order
parameter (black is for lower values, white for higher values). The
insets show the function $\mathcal{G}(m)$ in different areas of the
phase space. The dashed line represents the border between a one minimum
area in $\mathcal{G}$, and a two minima area. The solid line represents
a first order transition. The point where all the lines meet is the
analogue of a critical point.}
\end{figure}

Fig. \ref{fig:full_cusp} demonstrates the results of a numerical
calculation of the number of minima for $\mathcal{G}\left(a_{1},\, a_{2},\, m\right)$
around the critical point, and of the value of $m^{\star}$. At each
point $\left(a_{1},\, a_{2}\right)$ in the configuration space the
order parameter is the value of $m$ which minimizes $\mathcal{G}\left(a_{1},\, a_{2},\, m\right)$.
This value is obtained by minimizing $\mathcal{G}$ numerically. The
region in configuration space where two locally minimizing solutions
exist is also shown.

\subsubsection{Tricritical point analogue (butterfly catastrophe):}

We now move to look at profiles in the subspace of configurations
defined by 
\begin{equation}
\rho_{f}\left(x\right)=\frac{1}{2}-\delta+2\delta\cdot x+a_{1}\sin\left(\pi x\right)+a_{2}\sin\left(2\pi x\right)+a_{4}\sin\left(4\pi x\right).\label{eq:profile_sym_butterfly}
\end{equation}
In a manner similar to the one we used to find the cusp critical point,
we look at the first six coefficients of Eq. \ref{eq:G_expansion}
to find\numparts

\begin{eqnarray}
c_{1}= & a_{1}\,,\\
c_{2}= & \left(\delta-\pi a_{2}+2\pi a_{4}\right)\,,\\
c_{3}= & -\frac{\pi^{2}}{6}a_{1}\,,\\
c_{4}= & \frac{\pi^{3}}{3}\left(a_{2}-8a_{4}\right)\,,\\
c_{5}= & \frac{\pi^{4}}{120}a_{1}\,,\\
c_{6}= & \frac{2\pi^{5}}{45}\left(-a_{2}+32a_{4}\right)\,.
\end{eqnarray}
\endnumparts Higher order terms do not vanish. Using standard arguments
a proper choice of $c_{1},c_{2},c_{3},c_{4},c_{5},c_{6}$ which gives
$m^{\star}$ small, justifies the truncation of the series. It is
rather straightforward to check that these values correspond to a
realizable configuration where $0\leq\rho_{f}(x)\leq1$. Similar to
an expansion about a tricritical point (or a butterfly catastrophe)
we find 
\begin{eqnarray}
\mathcal{G}\left(m\right) & \simeq & 2\log\left(\frac{1+\delta}{1-\delta}\right)\cdot\left(c_{0}+a_{1}m+\left(\delta-\pi a_{2}+2\pi a_{4}\right)m^{2}-\frac{\pi^{2}}{6}a_{1}m^{3}\right.\label{eq:G_expansion_butterfly}\\
 &  & \left.+\frac{\pi^{3}}{3}\left(a_{2}-8a_{4}\right)m^{4}+\frac{2\pi^{5}}{45}\left(-a_{2}+32a_{4}\right)m^{6}\right)\;,\nonumber 
\end{eqnarray}
where the tricritical point is specified by 
\begin{equation}
\left(a_{1}^{\star},a_{2}^{\star},a_{4}^{\star}\right)=\left(0,\frac{4}{3}\frac{\delta}{\pi},\frac{\delta}{6\pi}\right)\,.
\end{equation}
Note that the coefficient of $m^{6}$ around the tricritical point
is positive. On the $a_{1}=0$ plane there is an analogue of a $\lambda$-line
\cite{chaikin2000principles}, with a second order phase transition
line connected to a first-order transition line at the tricritical
point. As we approach that point along the $a_{2}=8a_{4}$ line the
order parameter behaves as $m^{\star}\propto\left(\delta-\pi a_{2}+2\pi a_{4}\right)^{\nicefrac{1}{4}}$.

\begin{figure}[h]
\includegraphics[width=0.5\textwidth]{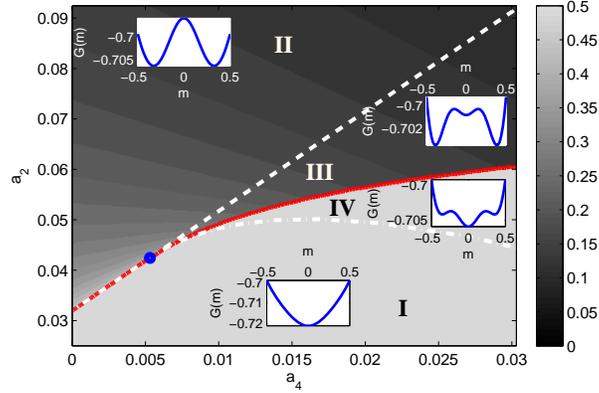}

\caption{\label{fig:full_sym_butterfly} The `phase diagram' for $\rho_{f}\left(x\right)=\rho_{0}+2\delta\cdot x+a_{2}\sin\left(2\pi x\right)+a_{4}\sin\left(4\pi x\right)$
with $\delta=0.1$. The background color represents the order parameter
(black is for lower values, gray is for higher values), where a lower
value was chosen where there were two competing values (due to the
symmetry). The phase space is divided into 4 regions. In region (I)
there is one minimum (corresponding to a symmetric history), in region
(II) there are two symmetric minima, and in regions (III) and (IV)
there are three different minima. The solid red line is a first order
transition line where three different minimizing solutions coexist.
The red-white line is a second order transition. The upper dashed
white line is a crossing between having one/three minima, and the
lower dashed white line is a crossing where one minimum transforms
to a maximum and two extra minima. The blue dot where all the lines
meet is a tricritical point . The insets show the function $\mathcal{G}(m)$
in different areas of the phase space.}
\end{figure}

Fig. \ref{fig:full_sym_butterfly} demonstrates the above structure
on the plane $a_{1}=0$. It is essentially a textbook tricritical
behavior in configuration space. The figure also shows the corresponding
$\mathcal{G}(m)$ at different locations in the plane.

\subsubsection{Multicritical points:}

With the above examples of singular behavior it is natural to ask
if configurations where more than three locally minimizing solutions,
which all give the same value of the LDF, exist. In analogy with the
above results this should imply the existence of analogues of multicritical
points of general order $s$. To identify such multicritical points
we look for configurations where the first $2s-1$ derivatives of
$\mathcal{G}(m)$ with respect to $m$ vanish. Close to such points
(where $2s-1$ coefficients disappear), it is safe to say that the
minimizing $m^{\star}$ is small, and so if the $2s$-th coefficient
is positive it will be dominant. Then terms in the Landau expansion
with $n>s$ can be neglected. In \ref{sec:appendix-coefficients}
we show that such a solution can be found, and that we need $n_{max}=s-1$
sine function in Eq. \ref{eq:profiles} to achieve that. The multicritical
point $\left(a_{1}^{\star},a_{2}^{\star},...,a_{2n_{max}}^{\star}\right)$
is found at 
\begin{equation}
a_{2n-1}^{\star}=0\,\label{eq:critical-point-odd}
\end{equation}
for the odd coefficients, while for the even coefficients we get 
\begin{equation}
a_{2n}^{\star}=\frac{2\left(\left(s-1\right)!\right)^{2}}{n\pi\left(s-1+n\right)!\left(s-1-n\right)!}\delta\,.\label{eq:critical-point-even}
\end{equation}
with $n$ between $1$ and $s-1$. For $c_{2s}$ we find 
\begin{equation}
c_{2s}=\frac{\left(-1\right)^{s+1}}{\left(2s\right)!}\sum_{n=1}^{s-1}\left(-1\right)^{n}\left(2n\pi\right)^{2s-1}a_{2n}^{\star}=\frac{\left(-1\right)^{s+1}}{\left(2s\right)!}\sum_{n=1}^{s-1}\left(-1\right)^{n}\frac{4\left(2n\pi\right)^{2s-2}\left(\left(s-1\right)!\right)^{2}}{\left(s-1+n\right)!\left(s-1-n\right)!}\delta\,.
\end{equation}
By substituting numerically for $s$ it can be seen that this expression
is positive for all $s$. This justifies the termination of the Landau
expansion at $c_{2s}$. One can easily check the validity of the expressions
by substituting $s=2$ for an Ising (cusp) singularity, and $s=3$
for a tricritical (butterfly) singularity, both of which were presented
above.

\begin{figure}
\includegraphics[width=0.5\textwidth]{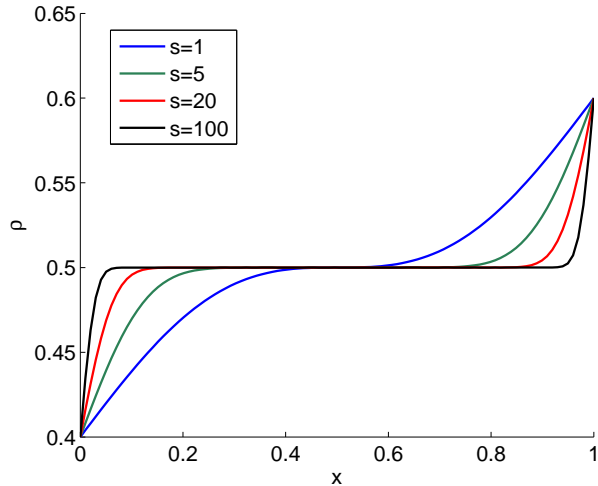}

\caption{\label{fig:critical_profiles}Plots of $\rho_{f}^{\star}\left(x\right)$
for different values of $s$. As $s$ increases, $\rho_{f}^{\star}$
is getting flatter. Each profile crosses the line $\rho=\nicefrac{1}{2}$
exactly once. A small deviation from these profiles will promptly
generate multiple extremal histories leading to these profiles.}
\end{figure}

Finally, to check that the coefficients $a_{n}$ correspond to realizable
configurations with $0\leq\rho(x)\leq1$ we plot the configurations
for different orders of $s$ in Fig. \ref{fig:critical_profiles}.
As can be seen all configurations are indeed realizable. In particular,
it is easy to see (using standard Fourier methods) that for $s\rightarrow\infty$,
the profile converges to: 
\begin{equation}
\rho_{f,\, s\rightarrow\infty}^{\star}\left(x\right)\longrightarrow\frac{\rho_{1}+\rho_{0}}{2}.\label{eq:critical_profile_inf}
\end{equation}

\section{Summary and discussion}

In this paper we focused on the structure of singularities in bulk
driven transport models. With a finite bulk field (a WASEP) we demonstrated
numerically that as the strength of the bulk bias increases analogues
of critical and tricritical point singularities appear in configuration
space. With the bulk drive infinite (a PASEP), we obtained an analytical
mapping between the large deviation functional and an effective Landau
theory. Remarkably, analogues of multicritical points of any order
have been identified. The direct mapping to a Landau theory is not
obvious. As discussed in \cite{bunin2013cusp} the applicability of
the Landau theory relies on the fact that the Hessian matrix, which
characterizes the stability of each history, has a single eigenvalue
which vanishes at the (multi)critical point. While here the measurement
of the Hessian proved too hard the exact mapping to a Landau theory
in the infinite field case as well as the numerics of the exponent
for a cusp singularity at small fields suggest that this is indeed
the case. It is an interesting questions to see if other models can
shown more complicated behaviors.

Finally, in the work presented above the Landau theory was proved
only in the infinite field limit and without directly relating the
order parameter to the histories. It would be interesting to see if
this can be done rigorously.

\textit{Acknowledgments:} We would like to thank Daniel Podolsky for
many useful comments and discussions. The work has been supported
by ISF and BSF grants.

\appendix

\section{\label{sec:appendix-shooting}Shooting method for more than one solution}

As discussed in the text, we are interested in detecting multiple
solutions of the ordinary differential equation of the mapping Eq.
\ref{eq:phi_to_rho_mapping}. To do this we first rewrite the equation
as a standard second order non-linear boundary value problem: 
\begin{eqnarray*}
\varphi_{xx} & =f\left(x,\varphi,\varphi_{x}\right)=\varphi_{x}\left(E-\varphi_{x}\right)\left(\frac{1}{1+e^{\varphi}}-\rho\right)\ ,
\end{eqnarray*}
\begin{equation}
\varphi\left(0\right)=\varphi_{0}=\log\left(\frac{\rho_{0}}{1-\rho_{0}}\right)\quad,\quad\varphi\left(1\right)=\varphi_{1}=\log\left(\frac{\rho_{1}}{1-\rho_{1}}\right)\ .
\end{equation}
Common algorithms for solving such problems, such as the shooting
or relaxation methods, usually look for \emph{one} solution \cite{press2007numerical}.
In the case of multiple solutions, the solution obtained by these
methods is dependent on the initial guess.

\begin{figure}
\includegraphics[width=0.5\textwidth]{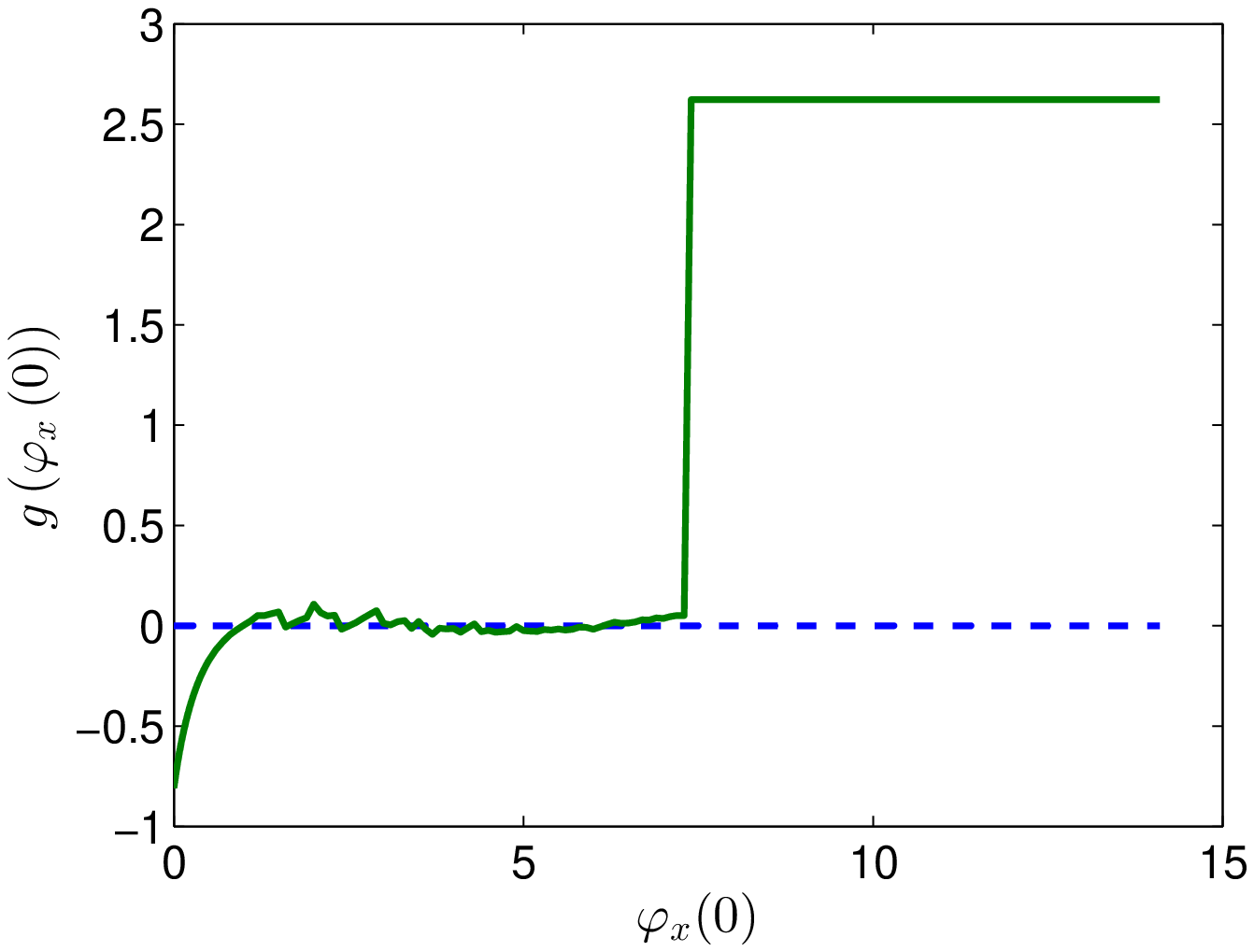}

\caption{\label{fig:g}An example of the function $g\left(\varphi_{x}\left(0\right)\right)$,
defined in Eq. \ref{eq:g}. Note how the function is very noisy and
could potentially yield excessive results if we do not double-check
to see that it really implies the existence of different solutions.
We used the fact that $\varphi$ is monotonic to stop integration
if the value $\varphi\left(c\right)$ at some point $c\in\left(0,1\right)$
exceeds the value of $\varphi_{1}$ by far. This is why the function
is clipped at the end.}
\end{figure}

To this end, in order to find all the possible solutions of the equation,
we used a modified approach, in which we use a range of possible values
for $\varphi_{x}\left(0\right)$. For each such value we integrate
the differential equation using standard initial value problem methods,
and look at the value $\varphi_{1}^{'}\left(\varphi_{x}\left(0\right)\right)$,
the value of $\varphi$ obtained at the right of the interval given
the initial value of $\varphi_{x}\left(0\right)$. We then look at
the function: 
\begin{equation}
g\left(\varphi_{x}\left(0\right)\right)=\varphi_{1}^{'}\left(\varphi_{x}\left(0\right)\right)-\varphi_{1}\ .\label{eq:g}
\end{equation}
Whenever $g\left(\varphi_{x}\left(0\right)\right)=0$, the solution
obtained by the integration of the differential equation for that
initial condition is a solution to the original boundary value problem.
The problem is now reduced to finding the roots of $g\left(x\right)$.
We do that by first calculating $g\left(x\right)$ on a coarse range
of values, find the intervals where we detect a crossing with zero.
Then we feed these intervals as initial intervals to a root finding
algorithm, such as Newton-Raphson. Since the function $g\left(x\right)$
is very noisy (see Fig. \ref{fig:g}), it may generate too many crossings
with zero, which might produce an excessive number of solutions.

To avoid over-counting the number of solutions, we put the solutions
obtained as an initial guess for a high accuracy boundary value problem
solver, on $L$ bins in the interval $\left[0,1\right]$. Obviously
the numerical accuracy of the results depends on the value of $L$.
Near a critical point, all the solutions are very close to each other,
so the ability to distinguish between the solutions is most vital
near the critical points (see main text). Typical values of $L$ used
were between $100$ and $5000$.

\section{\label{sec:appendix-coefficients}Derivation of coefficients for
the critical density profile at infinite bulk drive}

Here we derive the coefficients of the multicritical profile of order
$s$, namely Eqs. \ref{eq:critical-point-odd}, \ref{eq:critical-point-even}.
We start from Eqs. \ref{eq:multicritical_c_odd}, \ref{eq:multicritical_c_even}
for the coefficients. Our goal is to solve for the first $2s-1$ coefficients.
It is evident that $c_{n}$ with even and odd $n$ involve only even
and odd sine functions, respectively. We can use this fact and obtain
two sets of uncoupled equations for the coefficients of the multicritical
profile of order $s$. We denote this point by $\left(a_{1}^{*},a_{2}^{*},...,a_{s}^{*}\right)$.
For the odd coefficients we have the following set of equations\numparts\label{eq:most-critical-point}
\begin{equation}
A\cdot\boldsymbol{a}_{odd}^{*}=\left(\begin{array}{cccc}
1 & -1 & \dots & \left(-1\right)^{n+1}\\
-\pi^{2} & \left(3\pi\right)^{2} & \dots & \left(-1\right)^{n}\left(\left(2n-1\right)\pi\right)^{2}\\
\vdots &  & \ddots & \vdots\\
\left(-1\right)^{k}\left(\pi\right)^{2k-2} & \left(-1\right)^{k+1}\left(3\pi\right)^{2k-2} & \cdots & \left(-1\right)^{k+n}\left(\left(2n-1\right)\pi\right)^{2k}
\end{array}\right)\left(\begin{array}{c}
a_{1}^{*}\\
a_{3}^{*}\\
\vdots\\
a_{2s-1}^{*}
\end{array}\right)=\left(\begin{array}{c}
0\\
0\\
\vdots\\
0
\end{array}\right)\,,\label{eq:most_critical_point_odd}
\end{equation}
and for the even coefficients we have 
\begin{equation}
B\cdot\boldsymbol{a}_{even}^{*}=\left(\begin{array}{cccc}
-2\pi & 4\pi & \dots & \left(-1\right)^{n}2n\pi\\
\left(2\pi\right)^{3} & -\left(4\pi\right)^{3} & \dots & \left(-1\right)^{n+1}\left(2n\pi\right)^{3}\\
\vdots &  & \ddots & \vdots\\
\left(-1\right)^{k}\left(2\pi\right)^{2k-1} & \left(-1\right)^{k+1}\left(4\pi\right)^{2k-1} & \cdots & \left(-1\right)^{k+n}\left(2n\pi\right)^{2k-1}
\end{array}\right)\left(\begin{array}{c}
a_{2}^{*}\\
a_{4}^{*}\\
\vdots\\
a_{2s-2}^{*}
\end{array}\right)=\left(\begin{array}{c}
2\delta\\
0\\
\vdots\\
0
\end{array}\right)\,.\label{eq:most_critical_point_even}
\end{equation}
\endnumparts In both matrices we used the indices $k$ and $n$ to
enumerate the rows and columns, respectively. The matrices $A,\, B$
are both regular. To see that, we note that the determinant of the
matrix $A$ is equal up to a sign to that of the Vandermonde matrix
\cite{meyer2000matrix} 
\begin{equation}
\tilde{A}=\left(\begin{array}{cccc}
1 & 1 & \dots & 1\\
\pi^{2} & \left(3\pi\right)^{2} & \dots & \left(\left(2n-1\right)\pi\right)^{2}\\
\vdots &  & \ddots & \vdots\\
\left(\pi^{2}\right)^{k-1} & \left(\left(3\pi\right)^{2}\right)^{k-1} & \cdots & \left(\left(\left(2n-1\right)\pi\right)^{2}\right)^{k-1}
\end{array}\right)\,,
\end{equation}
which is regular. The determinant of the matrix $B$ is equal to that
of another matrix 
\begin{equation}
\tilde{B}=\left(\begin{array}{cccc}
-2\pi & 4\pi & \dots & \left(-1\right)^{n}2n\pi\\
-\left(2\pi\right)^{3} & \left(4\pi\right)^{3} & \dots & \left(-1\right)^{n}\left(2n\pi\right)^{3}\\
\vdots &  & \ddots & \vdots\\
-\left(2\pi\right)^{2k-1} & \left(4\pi\right)^{2k-1} & \cdots & \left(-1\right)^{n}\left(2n\pi\right)^{2k-1}
\end{array}\right)\,,
\end{equation}
which can be decomposed into two matrices, both of which are regular
\begin{equation}
\tilde{B}=\left(\begin{array}{cccc}
1 & 1 & \dots & 1\\
\left(2\pi\right)^{2} & \left(4\pi\right)^{2} & \dots & \left(2n\pi\right)^{2}\\
\vdots &  & \ddots & \vdots\\
\left(\left(2\pi\right)^{2}\right)^{k} & \left(\left(4\pi\right)^{2}\right)^{k} & \cdots & \left(\left(2n\pi\right)^{2}\right)^{k}
\end{array}\right)\cdot\left(\begin{array}{cccc}
-2\pi & 0 & \dots & 0\\
0 & 4\pi & \dots & 0\\
\vdots &  & \ddots & \vdots\\
0 & 0 & \cdots & \left(-1\right)^{n}2n\pi
\end{array}\right)\,.
\end{equation}
Since the matrix $\tilde{B}$ is composed of two regular matrices,
it is also regular, and so is the matrix $B$. It follows that the
set of Eqs. \ref{eq:most-critical-point} has a single solution. For
the odd part it is clear that the only solution is the trivial solution,
hence Eq. \ref{eq:critical-point-odd}. For the even part, Cramer's
rule \cite{cramer1750introduction} can be used to obtain a solution
to the linear set of equations, which yields Eq. \ref{eq:critical-point-even}.

\section*{References}

\bibliographystyle{iopart-num}
\bibliography{asep_singularities}

\end{document}